\documentclass[final,3p,twocolumn]{elsarticle}
\usepackage{graphicx}
\usepackage{amssymb}
\usepackage{amsmath}

\journal{Physics Letters B}

\begin{document}

\begin{frontmatter}

\title{Two-Proton Radioactivity with 2p halo in light mass nuclei A$=$18$-$34}
\author[a]{G. Saxena}
\author[b]{M. Kumawat}
\author[c]{M. Kaushik}
\author[b] {S. K. Jain}
\author[d]{Mamta Aggarwal\corref{cor1}}
\cortext[cor1]{corresponding authors: Mamta Aggarwal, mamta.a4@gmail.com}
\address[a]{Department of Physics, Government Women Engineering College, Ajmer-305002, India}
\address[b]{Department of Physics, School of Basic Sciences, Manipal Univ., Jaipur-303007, India}
\address[c]{Department of Physics, Shankara Institute of
Technology, Kukas, Jaipur-302028, India}
\address[d]{Department of Physics, University of Mumbai, Kalina Campus, Mumbai-400098, India}
\begin{abstract}
Two-proton radioactivity with 2p halo is reported theoretically in light mass nuclei A $=$ 18-34. We predict $^{19}$Mg, $^{22}$Si, $^{26}$S, $^{30}$Ar and $^{34}$Ca as promising candidates of ground state 2p-radioactivity with S$_{2p}$ $<$ 0 and S$_{p}$ $>$ 0. Observation of extended tail of spatial charge density distribution, larger charge radius and study of proton single particle states, Fermi energy and the wave functions indicate 2p halo like structure which supports direct 2p emission. The Coulomb and centrifugal barriers in experimentally identified 2p unbound $^{22}$Si show a quasi-bound state that ensures enough life time for such experimental probes. Our predictions are in good accord with experimental and other theoretical data available so far.
\end{abstract}


\end{frontmatter}
Two-proton radioactivity, a simultaneous decay of two protons from exotic nuclei near or beyond proton drip line \cite{Pftzner,BLANKBORGE,SON,WOODDAV}, is a recently discovered ~\cite{BLANKBORGE,BLAN,PFUTZR}, a relatively lesser known exotic nuclear decay mode that has attracted a lot of attention in the recent times. For a 2p-radioactive nucleus S$_{p}$ $>$ 0 and S$_{2p}$ $<$ 0 which means that 1p should not be an open decay channel \cite{goldansky} so that the two valence protons are emitted simultaneously where the role of pairing interaction becomes significant.  This valence 2p weakly bound system in the decay channel might form a halo \cite{xu1,LIN,XUXX,LIN1} like structure in the low density nuclear matter which is spatially far from the tightly bound core system and may lead to enhanced probability of direct 2p decay. The pairing studies are expected to give new insights into the role of 2p halo of the 2 valence proton system in the direct 2p emission. However 2p correlations \cite{taka,oishi} in 2p decay and its mechanisms are still not clear and require further theoretical and experimental investigations \cite{Pftzner}. \par
So far 2p-radioactivity has been observed in the decay of ground state $^{45}$Fe \cite{Pftzner,Giovinazzo}, $^{54}$Zn \cite{Blank,Ascher}, $^{48}$Ni \cite{Dossat,Pomorski}, $^{67}$Kr \cite{goigoux}, $^{19}$Mg \cite{mukha1}, $^{30}$Ar \cite{IMukhaar}, $^{31}$Ar \cite{Koldste,LIS} $^{20}$Mg \cite{Wallace,LUND,SUN}, and excited states $^{22}$Mg and $^{94}$Ag (for 1p and 2p decay) \cite{ma,mukha} in experimental and theoretical \cite{mapscr,aggarwal} works have provided useful insights into this exotic decay mode of proton rich nuclei. Recent experimental observations of (i) $\beta$-delayed two-proton emission in the decay of $^{22}$Si \cite{xu} (ii) 2p correlated emissions in $^{28,29}$S but not in  $^{27,28}$P which indicated \cite{xu1,LIN,XUXX} that the diproton correlations may result from proton configuration in the initial state like 2p halo rather than the deformed orbit (iii) and the speculation \cite{LIN1} that the 2p halo plays an important role in the diproton emission, lack a precise theoretical description that has invoked the present study. The objective in this letter is to search for 2p halo and 2p unbound nuclear systems in even Z$=$12 to 20 in a theoretical framework and present for the first time an analysis to probe the interdependence between the weakly bound 2p halo like structure and the direct 2p emission. \par
2p-radioactivity which is essentially the energetically allowed simultaneous two-proton emission is investigated by employing two simple and effective well established theoretical formalisms (i) Relativistic mean-field plus state dependent BCS (RMF+BCS) approach \cite{walecka,boguta,suga,ring,yadav,saxena,saxena1} and (ii) Macroscopic-microscopic approach (Mac-Mic) with Nilson Strutinsky (NS) prescription \cite{aggarwal,MAMPRC}. We calculate one- and two-proton separation energies of proton rich Mg, Si, S, Ar and Ca isotopes near proton drip-line and identify $^{19}$Mg, $^{22}$Si, $^{26}$S, $^{30}$Ar and $^{34}$Ca as potential candidates of 2p-radioactivity in agreement with the available experimental data \cite{mukha1,IMukhaar,xu} for $^{19}$Mg, $^{22}$Si and $^{30}$Ar. The structural evidences for the existence of halo are searched in the above identified 2p emitters within the RMF+BCS approach by investigating spatial density distribution, larger charge radius, wave functions and single particle energies that reveal a halo like structure of weakly bound protons which is the highlight of this work. Another important feature here is to look for a quasi bound state using RMF+BCS to provide finite life time to enable experimental study in 2p unbound  $^{22}$Si which has been recently identified a candidate of 2p-radioactivity experimentally \cite{xu1}. Although various theoretical approaches \cite{Ascher,mukha1,grigo,zhao,yeun,delion} have been used to study 2p unbound nuclei, the RMF+BCS approach together with a realistic mean-field has proved a very useful tool especially for drip line nuclei as shown in our earlier works \cite{saxena,saxena1,cjp} where it has been used to study magicity, weakly bound structure and halos. (Description of detailed theoretical formalisms that are adequately described in our earlier works, are avoided here. Kindly see references (RMF+BCS~ \cite{saxena,saxena1} and Mac-Mic~\cite{aggarwal,MAMPRC}) for details). Our calculations show consensus with experimental and other theoretical works and provide much needed insights on this subject. \par
\begin{figure}[htb]
\centering
\includegraphics[width=0.5\textwidth]{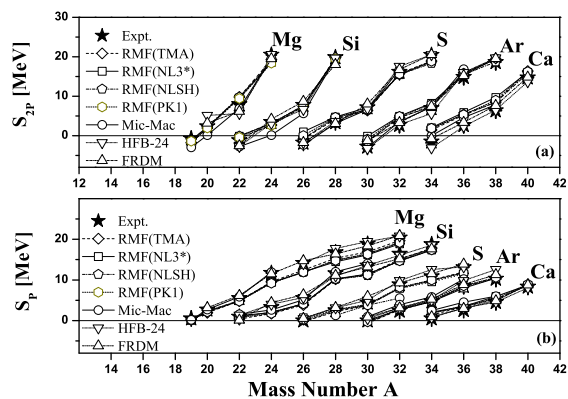}
\caption{Variation of(a) two- and (b) one- proton separation energy for Mg, Si, S, Ar and Ca isotopes along with available experimental data (with error bars) \cite{wang}, HFB \cite{goriely} and FRDM \cite{frdm2012} data.}
\label{fig1}
\end{figure}
Two- and one- proton separation energy (S$_{2p}$ and S$_{p}$) computed using RMF with TMA \cite{suga-tma}, NL3* \cite{nl3star}, NLSH \cite{ring3-nlsh} and PK1 \cite{pk1} parameters and Mac-Mic approach \cite{aggarwal} are shown in Fig. 1 (a) and (b) respectively where we find $^{19}$Mg, $^{22}$Si, $^{26}$S, $^{30}$Ar and $^{34}$Ca to be 2p unbound with S$_{2p}$ $<$ 0 and S$_{p}$ $>$ 0. Our calculated values of  S$_{2p}$ and S$_{p}$ and the prediction of the first two-proton unbound nuclei agree very well with the experimental data \cite{wang}, results from various parameters of RMF and other theories like nonrelativistic approach viz. Skyrme-Hartree-Fock method with the HFB-24 functional (described in Ref. \cite{goriely}) and FRDM \cite{frdm2012}. However, we find small anomaly for $^{34}$Ca and $^{26}$S for which all the theories and experiment show  S$_{2p}$ $<$ 0 whereas RMF shows $^{34}$Ca to be 2p bound with S$_{2p}$ $>$ 0 and one parameter (NL3*) of RMF shows $^{26}$S to be 2p bound with S$_{2p}$ $>$ 0. This anomaly requires further investigation. \par

To probe the structure of 2p unbound orbitals, we evaluate nuclear deformations using deformed RMF approach \cite{suga,yadav,singh,geng1,gambhir} and Mac-Mic approach ~\cite{aggarwal,MAMPRC} where $^{22}$Si and $^{34}$Ca are found to be spherical ($\beta$=0.0) in agreement with HFB \cite{goriely} and FRDM \cite{frdm2012} and $^{19}$Mg, $^{26}$S and $^{30}$Ar show large deformation which we study along with their neighbouring isotopes to check if the deformation shows any major structural changes in 2p unbound nuclei while moving from normally bound to unbound nuclei. Our estimated $\beta$ values for $^{19-21}$Mg (0.23, 0.19, 0.28), $^{26-28}$S (0.26, 0.26, 0.26) and $^{30-32}$Ar (0.22, 0.21, 0.20) show a gradual variation without any significant structural differences between deformed 2p emitters and their neighbouring isotopes. This suggests that the deformation appears to have no role in influencing 2p emission as speculated in  recent works \cite{xu1,LIN,XUXX}, where, mainly a 3-body simultaneous decay in phase space and 2p sequential emission for odd-Z $^{27}$P while a diproton-type decay for even-Z $^{28}$S indicated that the 2p halo-like structure rather than large deformation is responsible for 2p correlations which may play an important role in the mechanism of 2p emission. However there may not be any taboo to deformations influencing 2p emission in other nuclei which needs investigation in various mass regions theoretically as well as experimentally to reach any conclusion on this.\par
\begin{figure}[htb]
\centering
\includegraphics[width=0.5\textwidth]{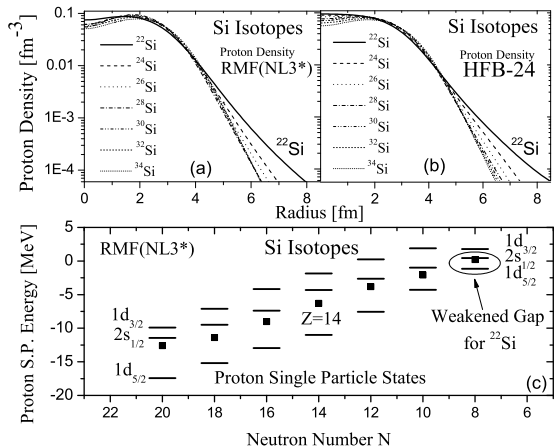}
\caption{Variation of charge density of $^{22-34}$Si vs. radius for (a) RMF(NL3*) (b) HFB \cite{goriely} (c) Variation of proton single particle states (1d$_{5/2}$, 2s$_{1/2}$, 1d$_{3/2}$) for $^{22-34}$Si using RMF(NL3*). Fermi energy is depicted by solid squares.}
\label{fig2}
\end{figure}
The 2p emission can happen either by sequential decay, direct diproton decay or 3-body democratic decay. However in view of the suggestion \cite{BLAN} that 2p halo would lead to larger spectroscopic factor for a direct 2p decay than for a sequential decay and the speculations \cite{xu1,LIN,XUXX,LIN1} that 2p halo plays an important role in the diproton emission, we look for the structural evidences for the existence and significance of 2p halo in 2p emitters. We study the charge density distribution and variation of proton single particle energies  of sd shell (1d$_{5/2}$, 2s$_{1/2}$, 1d$_{3/2}$) within the spherical framework of RMF with NL3* parameter for the experimentally identified 2p emitter $^{22}$Si \cite{xu}. From charge density distribution curves plotted in (Fig. 2(a)) it is evident that $^{22}$Si shows an extended tail in the charge density distribution which is much more extended than its neighboring isotope $^{24}$Si. This indicates weakly bound halo like structure of last two protons in the outer most shell in $^{22}$Si that has more probability of decay by 2p emission than relatively normally bound $^{24}$Si shows dependence of 2p halo on 2p emission. The charge densities of $^{22-34}$Si calculated by Skyrme-Hartree-Fock method with the HFB-24 functional are also shown in Fig. 2(b) where loosely bound structure of protons in $^{22}$Si match in a good deal with our calculations of RMF and is a strong evidence in support of decay via two-proton emission. Fig. 2(c) shows proton single particle energies  of sd shell where we note a large gap between 1d$_{5/2}$ and 2s$_{1/2}$ that produces a shell closure at Z$=$ 14 which is an area of recent interest as Si isotopes $^{34}$Si \cite{nature} and $^{48}$Si \cite{jia} have been identified as doubly magic nuclei in agreement with our calculations using RMF for $^{34}$Si (see in Fig. 2(c)). The gap between 1d$_{5/2}$ and 2s$_{1/2}$ in $^{34}$Si is found to be quite large around 6 MeV which reduces to 1.6 MeV for $^{22}$Si while moving from N$=$20 to N$=$8 towards the drip-line. This gap decreases from 7 MeV to 1.9 MeV and 7.4 MeV to 2.5 MeV using TMA and NLSH parameters respectively. Due to the reduced gap in $^{22}$Si, protons are partially occupied in higher state 2s$_{1/2}$ with  non$-$zero occupancy of 0.33 whereas the other isotopes $^{24-48}$Si have zero occupancy of proton in 2s$_{1/2}$ state. The Fermi energy (shown in Fig. 2(c)) in $^{22}$Si seems to be very close to this occupied 2s$_{1/2}$ state which indicates somewhat loosely bound valence protons near Fermi level in $^{22}$Si comprising a  2p halo like structure which is more likely to decay by 2p emission as compared to other isotopes of Si. This analysis provides a strong evidence towards the significant role of 2p halo in 2p decay.
\begin{figure}[htb]
\centering
\includegraphics[width=0.5\textwidth]{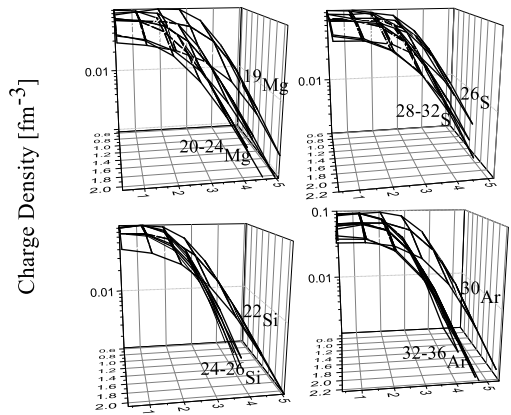}
\caption{Variation of charge density for few isotopes of Mg, Si, S and Ar near drip-line calculated by deformed RMF+BCS approach \cite{singh}.}
\label{fig3}
\end{figure}
The charge density for deformed two-proton emitters $^{19}$Mg, $^{26}$S, $^{30}$Ar along with $^{22}$Si calculated by RMF+BCS approach including deformation \cite{singh} is shown in Fig. 3. It is again remarkable to find the well spread density tail for these proton emitters in comparison to their neighboring isotopes indicating halo like structure. \par
Due to extended density distribution of loosely bound protons, the charge radii is expected to grow larger. In order to check this, we compute charge radii of drip-line isotopes of Mg, Si, S, Ar and Ca shown in Fig. 4 along with available experimental data \cite{angeli}. We observe much larger charge radii of $^{19}$Mg, $^{22}$Si, $^{26}$S, $^{30}$Ar (seen very evidently in Fig. 4) which is another strong evidence in the support of halo like structure in 2p emitters and thier possible influence on 2p decay which has not been shown earlier. Here we used deformed RMF approach with NL3*, TMA, NLSH and PK1 parameters that are in reasonable match with each other. Furthermore the difference between proton radius of $^{22}$Si and the daughter nucleus $^{20}$Mg is found very large which suggests that the removal of two loosely bound protons from $^{22}$Si after 2p emission results in a more confined configuration of $^{20}$Mg which is more stable with extra binding B/A = 6.724 MeV as compared to B/A = 6.058 MeV of $^{22}$Si. This shows that $^{22}$Si with spherical configuration is a loosely bound drip-line nucleus which decays via 2p emission and gains stability. Much larger charge radii in deformed 2p emitters $^{19}$Mg, $^{26}$S and $^{30}$Ar indicates loosely bound halo like structure in well deformed nuclei and is in conformity with the speculation \cite{xu1,LIN,XUXX,LIN1} that 2p halo is responsible for 2p correlations rather than the large deformation. However, the result for $^{34}$Ca is found albeit in difference for RMF theory where the charge radii is not much enlarged and also S$_{2p}$ $>$ $0$ although the experimental data \cite{wang} and other theories show S$_{2p}$ $<$0 and support energetically favoured 2p emission in $^{34}$Ca which needs further investigation.\par
\begin{figure}[htb]
\centering
\includegraphics[width=0.5\textwidth]{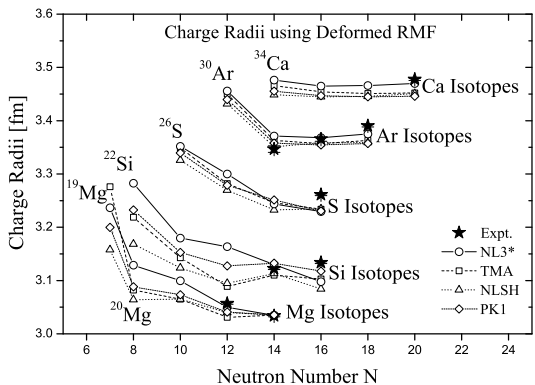}
\caption{Variation of charge radii for Mg, Si, S, Ar and Ca isotopes near proton drip line. Available experimental data (with error bars) is also shown \cite{angeli}.}
\label{fig4}
\end{figure}
Another remarkable feature of this work is to show the quasi bound state in two-proton emitter $^{22}$Si studied recently experimentally  \cite{xu} which indicates that it must have finite life time large enough to study such 2p unbound nucleus experimentally. In order to explain this, we evaluate and plot nuclear potential, centrifugal barrier calculated for d state (l$=$2), Coulomb barrier and total potential which is sum of all three in Fig. 5.  The nuclear potential along with Coulomb and centrifugal barriers has sufficient barrier height for the trapping of waves to form a quasi-bound state which enable $^{22}$Si to acquire rather long mean life time even it is unbound against two-proton emission and eventually $^{22}$Si may decay by the process wherein two-protons tunnel through the barrier leading to observation of two-protons radioactivity (shown by thick solid line in Fig. 5(a)). Due to the barrier effect, such a meta-stable state remains mainly confined to the region of the potential well and the wave function exhibits characteristics similar to that of a bound state. Indeed wavefunctions of proton valence shells 1d$_{5/2}$ and 2s$_{1/2}$ of $^{22}$Si are found similar to the wavefunctions of other bound isotopes (as seen in Fig. 5(b)) and remain in the potential region providing finite life time. For comparison we also show wavefunction of 2s$_{1/2}$ state of $^{20}$Si by solid line which significantly lies outside the potential region as seen in Fig. 5(b) that indicates an unbound nucleus. Interestingly, in $^{22}$O, the mirror nucleus of $^{22}$Si, Coulomb barrier is absent for N$=$14 therefore neutron 2s$_{1/2}$ state is found with -4.2 MeV energy that leads to a perfect bound nucleus.
\begin{figure}[htb]
\centering
\includegraphics[width=0.5\textwidth]{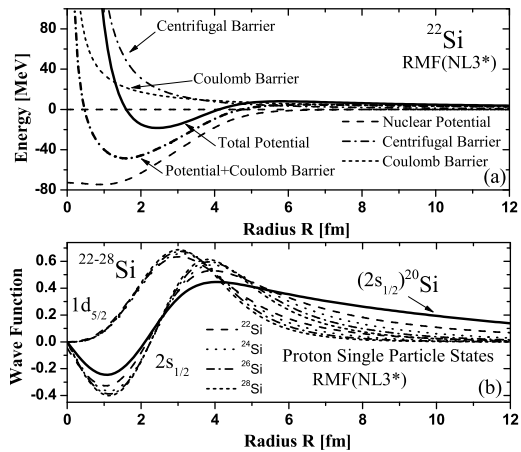}
\caption{(a) The RMF potential energy, centrifugal barrier energy for proton resonant states 1d$_{5/2}$ and Coulomb barrier as a function of radius for $^{22}$Si. Total potential (sum of potential energy, Centrifugal and Coulomb barrier), and sum of potential and Coulomb barrier are also shown. (b) Wavefunction of proton single particle 1d$_{5/2}$ and 2s$_{1/2}$ states are shown for $^{22-28}$Si. Wavefunction of proton 2s$_{1/2}$ state of $^{20}$Si is also shown.}
\label{fig5}
\end{figure}

To conclude, two-proton radioactivity, a relatively lesser known exotic nuclear decay mode along with  2p halo are investigated in light mass region around A $=$ 18-34 by employing (i) RMF+BCS approach and (ii) Macroscopic-microscopic approach with Nilson Strutinsky prescription. We predict $^{19}$Mg, $^{22}$Si, $^{26}$S, $^{30}$Ar and $^{34}$Ca as 2p-radioactive in their ground state with S$_{2p}$ $<$ 0 and S$_{p}$ $>$ 0. $^{22}$Si and $^{34}$Ca are found to be spherical whereas $^{19}$Mg, $^{26}$S and $^{30}$Ar show large deformations. Structural evidences for the existence of 2p halo in the 2p emitters are searched within the RMF+BCS approach. We observe long tail in extended spatial density distribution and larger charge radius in $^{19}$Mg, $^{22}$Si, $^{26}$S, $^{30}$Ar (except $^{34}$Ca) indicate 2p halo in 2p emitters. Study of single particle states reveals weakly bound halo like structure of protons due to weakening of shell gap and occupancy in proton 2s$_{1/2}$ state shown for experimentally identified nucleus $^{22}$Si further emphasizes the existence of 2p halo and its role in supporting 2p emission. The nuclear potential along with Coulomb and centrifugal barriers in $^{22}$Si appear to form a quasi-bound state to ensure a long delay in the decay of $^{22}$Si that may result its existence for finite life time and enable experimental probe. 2p-radioactivity and the link between 2p halo and diproton emission need more efforts on the experimental and theoretical fronts.\par

The authors thank Prof. H. L. Yadav, BHU, India, and Prof. L. S. Geng, Beihang University, China for valuable guidance. Authors GS and MA thank financial support by DST, India under YSS/2015/000952 and WOS-A respectively.

\end{document}